\documentstyle[11pt]{article}
 
\def\pa{\partial}
 
\def\g{\gamma} \def\G{\Gamma}

\def\e{\epsilon}

\def\l{\lambda} \def\L{\Lambda}
\def\m{\mu}

 \def\O{\Omega}

\def\ul{\underline}
\def\be{\begin{equation}}
\def\ee{\end{equation}}

\def\np{Nucl. Phys. }
\def\pl{Phys. Lett. }
\def\prl{Phys. Rev. Lett. }
\def\pr{Phys. Rev. }

\def\ijmp{Int. J. Mod. Phys. }

\begin{document}

\begin{flushright}

hep-th/9703176\\ \vspace{-.2in}
BRX TH--407, BROWN-HET-1078, US-FT-10/97\\

\end{flushright}

\vspace{-.4in}

\begin{center}
{\Large\bf A Note on the Picard--Fuchs Equations for\\
 N=2 Seiberg--Witten Theories}

Jos\'e M. Isidro,\footnote{Supported in part by Ministerio 
de Educaci\'{o}n y Ciencia, Spain.  Home address: 
Dep. de F\'{\i}sica de Part\'{\i}culas, Universidad 
de Santiago, 15706 Santiago, Spain.}$^{,4}$ Avijit 
Mukherjee,\footnote{Supported by the DOE under 
grant DE--FG02--92ER40706}
Jo\~{a}o P. Nunes,\footnote{Supported by the DOE 
under grant DE--FG02--91ER40688--Task A}
and 
Howard J. Schnitzer \footnote{Research supported 
in part by
the DOE under grant DE--FG02--92ER40706.}\\
Department of Physics,
Brandeis University, Waltham, MA 02254-9100, 
USA\footnote{isidro, mukherjee, 
schnitzer@binah.cc.brandeis.edu}\\
and\\
Department of Physics, Brown University, 
Providence, RI 02912, USA\footnote{nunes@het.brown.edu}\\
March 1997
\end{center}

\begin{quotation}
{\bf Abstract}:  A concise presentation of the PF
equations for $N$=2 Seiberg--Witten theories for the
classical groups of rank $r$ with $N_f$ massless
hypermultiplets in the fundamental representation is
provided.  For $N_f = 0$, all $r$ PF equations can be
given in a generic form.  For certain cases with
$N_f \neq 0$,  not all equations are generic. 
However, in all cases there are at least $r-2$
generic PF equations. For these cases the classical part
of the equations is generic, while the quantum part
can be formulated using a method described in a previous paper
by the authors, which is well suited to symbolic
computer calculations. 
\end{quotation}

\newpage

\noindent{\bf 1. ~Introduction}

Enormous advances have been made in 
understanding the low-energy properties 
of supersymmetric gauge theories \cite{001}.  In 
particular, the exact solution for the 
low-energy properties for the Coulomb phase 
of $N$=2 gauge theories with $N_f$ matter 
multiplets in the fundamental representation
is given in principle by a 
hyperelliptic curve \cite{001}-\cite{017}.  In fact a great 
deal of analysis is required to extract the 
strong-coupling physics from the curve characterizing 
the theory in question.  To do so, one defines 
the Seiberg--Witten (SW) period integrals
\renewcommand{\theequation}{1.\arabic{equation}}
\setcounter{equation}{0}
\be%1.1
\vec{\pi} = \left( 
\begin{array}{c}\vec{a}_D \\ \vec{a}
\end{array}
\right) \; ,
\ee
which are related to the prepotential ${\cal F} 
(\vec{a})$ characterizing the low-energy effective 
Lagrangian by
\be%1.2
a^i_D = \pa {\cal F}/\pa a_i \; . 
\ee
One  strategy for obtaining the necessary 
information is to derive a set of Picard--Fuchs 
(PF) equations for the SW period 
integrals.\footnote{
For alternate strategies see refs. \cite{018} and \cite{019}.}  The 
PF equations have been derived in a number of 
special cases for $N_f = 0$ and massless 
multiplets for $N_f \neq 0$ \cite{010,021,022,023,028}, 
using the strategy 
of Klemm, {\it et al.} \cite{002},
or Isidro {\it et al.} \cite{023}
[Examples of PF equations 
for massive multiplets are also known \cite{024,027,031}, but this 
situation will not be the concern of this paper.]  
A systematic method for finding PF equations, 
which is particularly convenient for symbolic 
computer computations, was given by Isidro, 
{\it et al.} \cite{023}.  Solutions to these
equations for low rank have also been considered 
\cite{010,021,025,026,027}, and studies of softly
broken $N$=2 theories \cite{020} have also been
carried out.

Although one can compute explicit PF equations 
for particular cases by the methods of Klemm,
{\it et al.}, \cite{002} or Isidro, {\it et al.}, 
\cite{023} it would 
be preferable to have an explicit, 
generic\footnote{By generic we mean PF equations which 
are applicable without specializing to any
particular rank $r$.  For example, see (2.17).
Sometimes we will refer to this as an analytic set of
PF equations.}
set of PF equations valid for each of the classical
groups, and those $N_f$ consistent with asymptotic 
freedom.  One expects $r$ 
PF equations, involving $r$ moduli for classical
groups of rank $r$.  A set of ($r-1$) equations
independent of the quantum scale  
for classical groups with $N_f = 0$ have been 
formulated by Ito and Sasakura \cite{022}, while a 
derivation of the remaining PF equation 
for $N_f =0$ for $A_r$, $B_r$ and $D_r$ 
was provided by Alishahiha \cite{028}.

It is the purpose of this paper
to extend these results so as to consider a 
complete set of $r$ PF equations for
the classical groups of rank $r$, with $N_f$ massless
hypermultiplets, with all equations
generic if possible.
The motivation for deriving such a set of equations 
is ultimately to extract general features of the 
low-energy strong-coupling behavior of $N$=2 
theories, without resorting to a case by case 
evaluation of the solutions of the equations.  
Generic solutions for the prepotentials in the 
\ul{weak-coupling} region have been given by D'Hoker, 
{\it et al}., \cite{018} and Matone, {\it et al}., 
\cite{019} but no comparable results are known 
for the strong-coupling regions, although we are 
presently considering such solutions to the 
equations presented in this paper.

In section 2, we review the PF equations available in
the literature \cite{022,028} for the case $N_f$=0,  
and derive analogous results for $C_r$.
However, these methods also require
some generalization to obtain the full set of PF
equations for all cases.

In section 3, we derive the $r$ PF equations for
$N_f \neq 0$, consistent with asymptotic freedom.
In certain cases, not all $r$ PF equations are
generic, although there are always at least $r-2$ generic
equations.  For the remaining
required PF equations we appeal to the $M$-matrix 
method of Isidro, {\it et al.}, to complete the 
discussion.  Thus, these equations must be dealt
with on a case by case basis. However, the $M$-matrix
methods are only required for the $\L$-dependent part 
of the additional PF equations, where $\L$ is the quantum scale of the
theory.

We note that this paper provides a concise
and complete presentation of the PF equations
presently available for
all classical groups, with
$N_f$ consistent with asymptotic freedom and with
$m=0$. We emphasize that it 
is not our purpose in this paper to
actually exhibit the explicit
computer calculations of the method of \cite{023},
whenever its use is required here.
Instead we wish to expose the maximal information
available in generic form in the PF equations.  It 
is our hope that this information will
facilitate the study of the Coulomb phase of $N=2$
theories in the strong-coupling region.

\noindent{\bf 2.~ PF Equation for $N_f = 0$}

\noindent A. ~\ul{Curves}

The hyperelliptic curves for the classical 
groups with $N_f = 0$ can be put in the following 
form
\renewcommand{\theequation}{2.\arabic{equation}}
\setcounter{equation}{0}
\be%2.1
y^2 = p^2 (x) - G(x)
\ee
with SW differential \cite{001,003,005,009,014}
\be%2.2
\l = \frac{x}{y} \left[
\frac{G^\prime}{G} \; \frac{p}{2} - p^\prime 
\right] dx \; .
\ee
In particular

\ul{$A_r = SU(r+1)$}  \cite{003,005}

\vspace{-.5in}

\begin{eqnarray}%2.3,4
p(x) & = & x^{r+1} - \sum^{r+1}_{i=2} \; 
u_i \: x^{r+1-i}\\
G(x) & = & \L^{2r+2}
\end{eqnarray}

\ul{$B_r = SO(2r+1)$}  \cite{006,009}

\vspace{-.5in}

\be %2.5
p(x)  =  P_r(x)\; ,
\ee
\be%2.6,7
\mbox{\rm where} \hspace{.4in} P_r(x) =  x^{2r} 
- \sum^r_{i=1} \; u_{2i} 
\: x^{2r-2i}
\ee
throughout this paper, and 
\be%2.7
G(x)  = x^2 \L^{4r-2}
\ee

\ul{$C_r = Sp(2r)$} \cite{014,022}

\vspace{-.5in}

\begin{eqnarray}%2.8,9
p(x) & = & x\: P_r (x) + \frac{1}{x} 
\: \L^{2r+2}\\
G(x) & = & \frac{1}{x^2} \L^{4r+4}
\end{eqnarray}
or equivalently
\be%2.10
y^2 = x^2 P^2_r (x) + 2P_r (x) \L^{2r+2}
\ee

\ul{$D_r = SO(2r)$} \cite{004,006}

\vspace{-.5in}

\begin{eqnarray}%2.11,12
p(x) & = & P_r(x)\\
G(x) & = & x^4 \L^{4r-4}
\end{eqnarray}

\noindent B. ~\ul{PF Equations}

Define $\pa/\pa u_i = \pa_i$ for convenience.

\ul{$A_r$}

\vspace{-.5in}

\begin{eqnarray}%2.13,14
\pa_i \l & = & - \frac{x^{r+1-i}}{y} \: 
dx + d (*)
\hspace{.4in} (i=2 \; \mbox{\rm to} \; r+1)
\\
\pa_i\pa_j \l & = & - \frac{x^{2r+2-i-j}}{y^3} 
\: p(x) dx + d (*)
\end{eqnarray}
which implies that ${\cal L}_{i,j;p,q} \;\; 
\vec{\pi} = 0$ where 
\be%2.15
{\cal L}_{i,j;p,q}= \pa_i\pa_j - \pa_p\pa_q
\ee
such that $i+j = p+q$.  Then
\be%2.16
\frac{d}{dx} \, \left( \frac{x^n}{y} \right) 
= n \, \frac{x^{n-1}}{y}
- \left( \frac{x^n p^\prime (x)}{y^3} \right) 
p(x) \; .
\ee
Using (2.3), (2.13) and (2.14) one obtains 
${\cal L}_n \;\;\vec{\pi} = 0$, where
\be
{\cal L}_0  = (r +1) \pa_2\pa_{r}
- \sum^{r+1}_{j=2} \, (r + 1-j) u_j\;\;
\pa_{j+1} \pa_{r+1}\nonumber \\
\ee
\begin{eqnarray}%2.17
{\cal L}_n & = & -n \: \pa_{r+2-n}
+ (r +1) \pa_2\pa_{r-n} \nonumber \\
&&  - \sum^{r+1}_{j=2} \, (r + 1-j) u_j\;\;
\pa_{r+2-n} \pa_j
\hspace{.4in} (n=1 \; \mbox{\rm to} \; r-2)
\end{eqnarray}
Equations (2.17) and (2.18) provide $r-1$ PF
equations \cite{022}.  The additional PF equation
follows from the strategy laid out in equations 
(10) and (11) of ref. \cite{028}, and our 
generalization thereof.

{}From (2.16) one may write
\begin{eqnarray}%2.18
D & = & - (r+1) d \left( \frac{x^{r+2}}
{y} \right) +
        \sum^{r+1}_{j=2} \: (r+1+j) u_j\: d 
\left( \frac{x^{r+2-j}}{y}            
\right) \nonumber \\[.1in]
& = & \l - \left[ \: \sum^{r+1}_{j=2} \: j(j-2) u_j \: 
        \frac{x^{r+1-j}}{y} \; \right. 
\nonumber \\[.1in]
& & + \; \sum^{r+1}_{i,j=2} \: ij\, u_i\,u_j \:
      \frac{x^{2r+2-i-j}}{y^3} \: p(x) 
\nonumber \\[.1in]
& & - \; \left. (r+1)^2 \: \frac{\L^{2r+2}}{y^3} 
\: p(x)\, \right] \; dx \; .
\end{eqnarray}
Using equations (2.13) and (2.14), this leads to
\begin{eqnarray}%2.19
{\cal L}_{r-1} & = & 1 + \sum^{r+1}_{j=2} \: 
j (j-2) u_j \, \pa_j +
\sum^{r+1}_{i,j=2} \: ij \, u_i \, u_j \, 
\pa_i \, \pa_j \nonumber \\
&& - (r + 1)^2 \, \L^{2r+2} \: \pa^2_{r+1} \; ,
\end{eqnarray}
which was conjectured in equation (68) of ref. 
\cite{022} and given in ref. \cite{028}.

\ul{$B_r$}
 
\vspace{-.5in}
\begin{eqnarray}%2.20,21
\pa_{2i} \, \l & = & \frac{-x^{2r-2i}}{y} 
\: dx + d(*)\\[.1in]
\pa_{2i}\pa_{2j} \, \l & = & \frac{-x^{4r-2i-2j}}
{y^3} \: p(x) dx + d(*)
\end{eqnarray}
which implies ${\cal L}_{2i,2j; 2p,2q} 
\vec{\pi} = 0$ where
\be% 2.22
{\cal L}_{2i,2j; 2p,2q} = \pa_{2i}\pa_{2j} 
- \pa_{2p}\pa_{2q}
\ee
with $i+j = p + q$.  Also
\be%2.23
\frac{d}{dx}\; \left( \frac{x^{2n+1}}{y} 
\right) = \frac{2nx^{2n}}{y} - 
\frac{[x^{2n+1}p^\prime (x) - x^{2n}p(x)]}
{y^3} \; p(x) \; .
\ee
{}From (2.21) and (2.22), one has
\begin{eqnarray}%2.24
{\cal L}_n & = & 2n\pa_{2r-2n} - (2r-1) 
\pa_2\pa_{2r-2n-2} \nonumber \\
&& + \sum^r_{j=1} (2r-2j-1) u_{2j} \pa_{2j} 
\pa_{2r-2n} \hspace{.4in}
(n=0 \;\mbox{\rm to} \; r-2)
\end{eqnarray}
{}From (2.24) we have 
\begin{eqnarray}%2.25
D & = & -(2r-1) \, d\left( \frac{x^{2r+1}}
{y} \right)\nonumber \\[.1in]
&& + \; \sum^r_{j=1} (2r + 2j -1 ) u_{2j} d
\left( \frac{x^{2r+1-2j}}{y} \right) 
\nonumber \\[.1in]
& = & \l - \left[ 4 \sum^r_{j=1} j (j-1) u_{2j} 
\, \frac{x^{2r-2j}}{y} \right.
\nonumber \\ [.1in]
&& + \; 4 \sum^r_{i,j=1} ij \, u_{2i} u_{2j} 
\, \frac{x^{4r-2i-2j}}{y^3}\:
p(x) \nonumber \\
&& - \; \left. (2r-1)^2 \: \L^{4r-2} \, 
\frac{x^2}{y^3} \, p(x) \right] dx \; .
\end{eqnarray}
Using (2.21) and (2.22), this gives
\begin{eqnarray}%2.26
{\cal L}_{r-1} & = & 1 + 4 \sum^r_{j=1} 
\: j (j-1) u_{2j} \, \pa_{2j}
\nonumber \\[.1in]
&& + \; 4 \sum^r_{i,j=1}  ij \, u_{2i} \, 
u_{2j} \, \pa_{2i} \, \pa_{2j} 
\nonumber \\[.1in]
&& - (2r - 1)^2 \, \L^{4r-2} \: \pa_{2r}
\pa_{2r-2} \; .
\end{eqnarray}

\ul{$C_r$}
 
\vspace{-.5in}
\begin{eqnarray}%2.27,28
\pa_{2i} \, \l & = & \frac{-x^{2r-2i+1}}
{y} \: dx + d(*)  \\[.1in]
\pa_{2i}\pa_{2j} \, \l & = & \frac{-x^{4r-2i-2j+2}}
{y^3} \: p(x) dx + d(*)
\end{eqnarray}
so that (2.23) is valid for $C_r$.
\be%2.29
\frac{d}{dx} \, \left( \frac{x^{2n+2}}{y}
\right) =
\frac{(2n+3)x^{2n+1}}{y} - \frac{[x^{2n+2}
p^\prime (x) + x^{2n+1} p(x)]}{y^3} \; p(x) \; .
\ee
Thus
\begin{eqnarray}%2.30
{\cal L}_n & = & (2n+3) \pa_{2r-2n} - (2r+2) 
\pa_2\pa_{2r-2n-2} \nonumber \\
&& + \; \sum^r_{j=1} \, (2r + 2 -2j) u_{2j}
\pa_{2j} \pa_{2r-2n} \hspace{.4in}
(n = 0 \; \mbox{\rm to} \; r-2)
\end{eqnarray}
{}From (2.30) one can write
\begin{eqnarray}%2.31
D & = & -(2r+2) \, d\left( \frac{x^{2r+2}}
{y}\right) \nonumber \\[.1in]
&& + \; \sum^r_{j=1} (2r + 2 + 2j) u_{2j} 
\; d \left( \frac{x^{2r+2-2j}}{y} \right)
 \nonumber \\[.1in]
&& - \; (2r+2)^2 \; \L^{2r+2} \: d(1/y) 
\nonumber \\[.1in]
& = & \l - \left[ 4 \sum^r_{j=1} j (j-1) 
u_{2j} \, \frac{x^{2r-2j+1}}{y} \, 
+ 4 \sum^r_{i,j=1} \, ij \, u_{2i}u_{2j} \:
\frac{x^{4r-2i-2j+2}}{y^3}\: p(x) \right.
\nonumber \\ [.1in]
&& + \;  (2r+2)^2 \; \L^{2r+2} 
\sum^r_{j=1} 
(2r-2j) \, u_{2j} \; 
\frac{x^{2r-2j}}{y^3}\:
p(x) \nonumber \\
& - &  \left.  2r(2r+2)^2 \L^{2r+2} x^{2r} \, 
\frac{p(x)}{y^3} \right]\, dx
\end{eqnarray}
[It should be noted that equation (2.32) is a
generalization of the method of ref. 
\cite{028}.]  {}From (2.28) and (2.29)
\begin{eqnarray}%2.32
{\cal L}_{r-1} & = & 1 + 4 \sum^r_{j=1} \, 
j(j-1) u_{2j}\pa_{2j} \nonumber \\[.1in]
&& + \; 4 \sum^r_{i,j=1} \, ij\: u_{2i}
u_{2j}\pa_{2i}\pa_{2j} \nonumber \\[.1in]
&& + \;(2r+2)^2 \, \L^{2r+2} \, 
\sum^r_{j=1} (2r-2j) \, u_{2j}  \pa_{2j+2}
\pa_{2r} \nonumber \\
&& - \; 2r (2r +2)^2 \, \L^{2r+2} \pa_2 \pa_{2r}
\; .
\end{eqnarray}

\ul{$D_r$}
 
\vspace{-.5in}
\begin{eqnarray}%2.33,34
\pa_{2i} \, \l & = & \frac{-x^{2r-2i}}{y} 
\: dx + d(*)  \\[.1in]
\pa_{2i}\pa_{2j} \, \l & = & 
\frac{-x^{4r-2i-2j}}{y^3} \: p(x) dx + d(*)
\end{eqnarray}
and (2.23) is valid for $D_r$ as well. We have
\be%2.35
\frac{d}{dx}\; \left( \frac{x^{2n+1}}{y} \right) 
= (2n-1) \frac{x^{2n}}{y} - 
\frac{[x^{2n+1}p^\prime (x) - 2x^{2n}p(x)]}
{y^3} \; p(x) \; .
\ee
which from (2.34), (2.35) gives
\begin{eqnarray}%2.36
{\cal L}_n & = & (2n-1)\pa_{2r-2n} - (2r-2) 
\pa_2\pa_{2r-2n-2} \nonumber \\
&& + \sum^r_{j=1} (2r-2-2j) u_{2j} \pa_{2j} 
\pa_{2r-2n} \hspace{.4in}
(n=0 \;\mbox{\rm to} \; r-2)
\end{eqnarray}
{}From (2.36) 
\begin{eqnarray}%2.37
D & = & -(2r-2) \, d\left( \frac{x^{2r+1}}
{y} \right) + 2 \sum^r_{j=1} (r  -1 +j) 
u_{2j} d
\left( \frac{x^{2r+1-2j}}{y} \right) 
\nonumber \\[.1in]
& = & \l - \left[ 4 \sum^r_{j=1} j (j-1) u_{2j} 
\, \frac{x^{2r-2j}}{y} \right.
 + \; 4 \sum^r_{i,j=1} ij \, u_{2i} u_{2j} 
\, \frac{x^{4r-2i-2j}}{y^3}\:
p(x) \nonumber \\
&& - \; \left. (2r-2)^2 \: \L^{4r-4} \, 
\frac{x^4}{y^3} \, p(x) \right] dx \; ,
\end{eqnarray}
which, when combined with (2.34) and (2.35), 
provides the remaining equation
\begin{eqnarray}%2.38
{\cal L}_{r-1} & = & 1 + 4 \sum^r_{j=1} \, 
j(j-1) u_{2j}\pa_{2j} \nonumber \\[.1in]
&& + \; 4 \sum^r_{i,j=1} \, ij\: u_{2i}
u_{2j}\pa_{2i}\pa_{2j} \nonumber \\[.1in]
&& - \; (2r-2)^2 \, \L^{4r-4} \, 
\pa^2_{2r-2}
\end{eqnarray}

In this section we have presented the $r$ PF 
equations for $A_r$, $B_r$, $C_r$, and $D_r$
with $N_f = 0$ in analytic form.  Equations
(2.31)--(2.33) for $C_r$ are new.

\noindent C. ~\ul{Alternate Curves}

It is understood that the particular form of 
the hyperelliptic curve for a given theory is 
not unique.   A number of different curves might 
represent the same physics, perhaps with a 
redefinition of moduli in such a way that 
different definitions of moduli agree in the 
semi-classical region, though in the 
strong-coupling region they may  
differ considerably.  The point we make is that 
not all versions of curves for 
a given theory lead to a complete set of 
PF equations in analytic form. 
We illustrate this issue with a 
discussion of $C_r$ with $N_f = 0$. The version of 
the curve we use in (2.8)--(2.10) 
is from Ito and Sasakura \cite{022}.  
We observe from (2.31) 
and (2.33) that a complete set of $r$ PF equations 
in analytic form are available for this case.  
However, we can also consider the curve for $C_r$ 
used by D'Hoker, {\it et al.}, \cite{018}
\be%2.39
y^2 = p^2(x) - \L^{4r+4} 
\ee
where $p(x) = x^2 P_r(x)$.
Then defining, analogous to (2.32),
\be%2.38
D = -(2r+2) d \left( \frac{x^{2r+3}}{y} \right) +
\sum^r_{j=1} (2r +2+2j) u_{2j} d
\left( \frac{x^{2r+3-2j}}{y} \right) \; ,
\ee
one finds that this cannot be related
to $\l$ so as to provide a generic PF equation.

The lesson is that not all distinct versions
of a curve representing the same prepotential lead
to a complete set of $r$ PF equations in generic
form.  For our purposes (2.10) is preferred.

\noindent{\bf 3. ~PF Equations for $N_f \neq 0$}

\renewcommand{\theequation}{3.\arabic{equation}}
\setcounter{equation}{0}

\noindent A. ~\ul{Curves}

The hyperelliptic curves for the classical 
groups with $N_f \neq 0$ are also described 
by equations (2.1) and (2.2).  In particular

\ul{$A_r; \; 1 \leq N_f < (r+1)$} \cite{005}

\vspace{-.5in}

\begin{eqnarray}%3.1,2
p(x) & = & x^{r+1} - \sum^{r+1}_{i=2} u_i \, 
x^{r+1-i} \\
G(x) & = &  \L^{2r +2 - N_f} \: x^{N_f}
\end{eqnarray}

\ul{$A_r; \; (r+1) \leq N_f < (2r+2)$}\footnote{The
coefficient 1/4 in (3.3) is subject to testing by
means of microscopic instanton calculations.  In fact,
for $SU(3)$ with $N_f = 3,4$, and 5, this coefficient
is actually 1/16, 5/32, and 17/64 respectively 
\cite{030}.  We keep the coefficient 1/4 in (3.3)
and (3.22) {\it ff}, recognizing that this coefficient
is readily replaced as more information becomes
available. The same issue applies to the curves corresponding to other  
gauge groups for large $N_f$.} \cite{005}

\vspace{-.5in}

\begin{eqnarray}%3.3
p(x) & = & x^{r+1} - \sum^{r+1}_{i=2} u_i \, 
x^{r+1-i} \nonumber \\
&& + \; \frac{1}{4} \: \L^{2r +2 - N_f} \: 
x^{N_f-r-1}
\end{eqnarray}

and $G(x)$ as in (3.2).

\ul{$B_r; \; 1 \leq N_f < (r-1)$} \cite{006,009}

\vspace{-.5in}

\begin{eqnarray}%3.4,5
p(x) & = & P_r(x) \\
G(x) & = & \L^{4r-2-2N_f} \: x^{2+2N_f}
\end{eqnarray}

\ul{$B_r; \; (r-1) \leq  N_f < (2r-1)$} \cite{006,009}

\vspace{-.5in}

\begin{eqnarray}%3.6
p(x) & = & P_r(x) + \L^{4r-2-2N_f} 
x^{2N_f - 2r +2}\nonumber \\
G(x) & = & \L^{4r-2-2N_f} \: x^{2+2N_f}
\end{eqnarray}

\ul{$C_r; \; 1 \leq  N_f < 2r+2$} \cite{014,022}

\vspace{-.5in}

\begin{eqnarray}%3.7,8
p(x) & = & x \: P_r(x)\\
G(x) & = & \L^{4r+4-2N_f} \: x^{2N_f-2}
\end{eqnarray}

\ul{$D_r; \; 1 \leq N_f < (r-2)$} \cite{004,006}

\vspace{-.5in}

\begin{eqnarray}%3.9,10
p(x) & = & P_r(x) \\
G(x) & = & \L^{4r-4-2N_f} \: x^{4+2N_f}
\end{eqnarray}

\ul{$D_r; \; (r-2) \leq N_f < (2r-2)$} \cite{004,006}

\vspace{-.5in}

\begin{eqnarray}%3.11
p(x) & = & P_r(x)+ \L^{4r-4-2N_f} \: 
x^{2N_f-2r + 4} \nonumber\\
G(x) & = & \L^{4r-4-2N_f} \: x^{4+2N_f}
\end{eqnarray}

\noindent B. ~\ul{PF Equations}

\ul{$A_r: \; 1 \leq N_f < (r+1)$}

{}From
\begin{eqnarray}%3.12
\frac{d}{dx} \, \left( \frac{x^n}{y} \right) & = &
\left( n - \frac{N_f}{2} \right) \, \frac{x^{n-1}}{y}
\nonumber \\
&& - \; \left( r+1 - \frac{N_f}{2} \right) \,
\frac{x^{r+n}}{y^3} \: p(x) \nonumber \\
&& + \; \sum^{r+1}_{j=2} \, \left(
r+1-j-\frac{N_f}{2} \right) \, 
\frac{u_j \: x^{r+n-j} \, p(x)}{y^3}
\end{eqnarray}
and equations (2.13) and (2.14), one has
\begin{eqnarray}%3.13
{\cal L}_n & = & - \left( n - \frac{N_f}{2} \right) 
\pa_{r+2-n}
+ \left( r +1 - \frac{N_f}{2} \right)
\pa_2 \pa_{r-n}
\nonumber \\
&& - \; \sum^{r+1}_{j=2} \, \left(
r+1-j-\frac{N_f}{2} \right) \, u_j \, \pa_{r+2-n} 
\, \pa_j \; ,
\end{eqnarray}
valid for $1 \leq n \leq r-2$.  Although (3.12) is
applicable for $n=0$, some work will be required
to obtain an equation in terms of $\l$.

{}From  Eq. (3.12), with $n=0$ 
\begin{eqnarray}%3.14
\frac{d}{dx} \left( \frac{1}{y} \right)\: = & - &
\frac{N_f}{2} \left( \frac{1/x}{y} \right) 
\nonumber \\
& - & \left( r+1 - \frac{N_f}{2} \right) \,
\frac{x^r}{y^3} \: p(x) \nonumber \\
& + & \sum^{r+1}_{j=2} \, 
\left( r + 1 - j - \, \frac{N_f}{2} \right) \,
u_j \, \frac{x^{r-j}}{y^3} \: p(x)
\end{eqnarray}
It follows from the curve (3.1) that
\begin{eqnarray}%3.15
\frac{d}{dx} \left( \frac{1}{y} \right)\; = & - &
(r+1) \, \frac{x^r}{y^3} \, p(x) \nonumber \\
& + & \sum^{r+1}_{j=2} \,
(r+1-j) u_j \: \frac{x^{r-j}}{y^3} \:
p(x) \nonumber \\
& + & \frac{N_f}{2} \: \L^{2r+2-N_f} \:
\frac{x^{N_f-1}}{y^3} \; .
\end{eqnarray}
Using (2.13) and (2.14)
\begin{eqnarray}%3.16
d \, \left( \frac{1}{y} \right) & = &
(r+1) \pa_2\pa_r \; \l \nonumber \\
& - & \sum^{r+1}_{j=2} \, (r+1-j)
u_j\, \pa_{j+1}\pa_{r+1}\; \l 
\nonumber \\
& + & \frac{N_f}{2} \: \L^{2r+2-N_f} \:
\frac{x^{N_f-1}}{y^3}\: dx \; .
\end{eqnarray}
It is very satisfying that (3.16) coincides with
(2.17) in the limits $N_f = 0$ or $\L =0$,
as it must.  However, the term $x^{N_f-1}/y^3$
is not immediately expressible in terms of
$\l$, as one cannot apply (2.13) or (2.14) directly.
Instead we appeal to the  method of
Isidro, {\it et al.} \cite{023,029,031}.

Define, following equation (2.9) of ref. \cite{023},
\be%3.17
\O^{(\m )}_m = (-1)^{\m +1} \: 
\G (\m +1) \, \int \: \frac{x^m}{y^{2(\m +1)}} 
\: dx
\ee
where we have suppressed the dependence of 
$\O^{(\m )}_m$ on the moduli and the fixed 1-cycle
$\g$.  Thus, we are interested in relating
$\O^{(1/2)}_{N_f-1}$ to $\O^{(-1/2)}_m$.
One can relate $\O^{(1/2)}_{m^\prime}$ to
$\O^{(-1/2)}_m$ for values of $m$ and $m^\prime$
belonging to the basic range $R$: ($m=0$ to
$r-1$; $r+1$ to $2r$).  Since 
$\O^{(1/2)}_{N_f-1}$ belongs to the basic range 
for $1 \leq N_f \leq r$, one may use Eq. (2.19)
of Isidro {\it et al.}, \cite{023} to write the
matrix equation
\be%3.18
\O^{(-1/2)} = M \cdot \O^{(1/2)} \; .
\ee
Thus $\int \, dx \, \frac{x^{N_f-1}}{y^3}$
is related to the periods
$\int \, dx \, \frac{x^m}{y}$;
$m \: \e \: R$, in terms of the coefficient matrix
$M$, which is given in terms of polynomials
in the moduli and the quantum scale $\L$.

Note that the period is not yet given in terms 
of $\l$.  To complete this task, one may
follow the method of Sec. 3.3 of Isidro,
{\it et al.} \cite{023}.   We do not give the details 
here, but sketch the main ideas.  Equation
(3.18) relates $\O^{(1/2)}_{N_f-1}$
to the periods $\O^{(-1/2)}_m$, $m \: \e \: R$.
Then from (2.13)
\be%3.19
\O^{(-1/2)}_m = -{\rm i}\, \Gamma(1/2)\,\pa_{r+1-m} \, \int \, dx \, \l
\hspace{.4in} \mbox{\rm for} \hspace{.2in}
 m=0 \hspace{.1in} \mbox{\rm to} \hspace{.1in} 
r-1 \; .
\ee
However, if $m = r+1$ to $2r$, we need the
analogue of (3.9) of Isidro, {\it et al.} 
\cite{023}. In
our notation, schematically
\begin{equation}%3.20
\left(
\begin{array}{c}
\Omega^{(-1/2)}_{r+1}\vspace{-.15in} \\ 
\vdots \vspace{-.2in} \\ \Omega^{(-1/2)}_{2r}
\end{array}
\right) = -B^{-1}_i 
\left( \frac{\partial}{\partial u_i}
+ A_i \right) \: 
\left( 
\begin{array}{c}
\Omega^{(-1/2)}_0 \vspace{-.15in}\\
\vdots \vspace{-.2in}\\ \Omega^{(-1/2)}_{r-1}
\end{array}
\right)_ {\hspace{.4in}\textstyle(i =2 \hspace{.05in} 
\makebox{\rm to} \hspace{.05in} r+1)}      \; 
\end{equation}
[The matrices $A_i$ and $B_i$ depend on the moduli
and the quantum scale $\L$. No summation over $i$ is implied in 
the above equation. It  holds for any $i$ from 2 to $r+1$, 
and any convenient value of $i$ is sufficient for our purposes.]
Finally, one uses (3.19) in (3.20) to
express the periods in terms of $\int \l$.  
Clearly, the last term in (3.16) leads to
first and second order modular derivatives
of $\l$.  Thus (3.16) gives a second-order
PF equation.

The steps (3.18) to (3.20) do not have a
generic solution, although the iterative algorithms
of Isidro, {\it et al.}, are well suited to
symbolic computer calculations.  They
must be dealt with on a case by case basis.  However,
observe that the computer calculations are only
needed for the $\L$-dependent part of (3.16).

We need one more PF equation. Using the curve 
(3.1), (3.2), the generalization
of (2.19) and (2.20) gives us
\begin{eqnarray}%3.21
{\cal L}_{r-1} & = &
1 +  \: \sum^{r+1}_{j=2} \, j(j-2) u_j\pa_j
\nonumber \\
& + &  \sum^{r+1}_{i,j=2} \,i j u_i u_j
\pa_i \pa_j\nonumber \\
& - & 
\left[ \left( r+1 -\frac{N_f}{2} \right)^2 
\L^{2r+2-N_f}
\pa_i \pa_j \right]_{i+j = 2r+2 -N_f}
\end{eqnarray}
which completes the set of $r$ PF equations.

\ul{$A_r: \;  (r+1) \leq N_f \leq 2r+1$}

The curve given by (3.3) allows the replacement
\begin{eqnarray}%3.22
u_{2r+2-N_f} & \longrightarrow & 
\bar{u}_{2r+2-N_f} \nonumber \\
& = & u_{2r+2-N_f} - \frac{1}{4} 
\L^{2r+2-N_f} \; ,
\end{eqnarray} 
for $r+1 \leq N_f \leq 2r$.
Thus, (3.13) gives $r-2$ PF equations,
with the replacement (3.22) understood for 
these flavors.

For $N_f = 2r+1$, (3.22) does not describe
an allowable shift.  Nonetheless for this case
one easily finds instead of (3.13)
\begin{eqnarray}%3.23
{\cal L}_n & = & - \left( n - r -\frac{1}{2} 
\right) \pa_{r+2-n} + \frac{1}{2} \:
\pa_2 \pa_{r-n} \nonumber \\
& - & \sum^{r+1}_{j=2} \left( \frac{1}{2}
-j \right) u_j \, \pa_{r+2-n} \pa_j \nonumber \\
& + & \frac{1}{8} \, \L \, \pa_{r+1-n} \: \pa_2 \; ,
\end{eqnarray}
valid for $1 \leq n \leq r-2$.  [One can understand
the additional term in (3.23) as the formal shift
$u_1 \rightarrow \bar{u}_1 = u_1 - \frac{1}{4} \L$,
with $u_1$ set to zero at the end.]

Since the relevant curve is (3.3), the
equation ${\cal L}_0$ differs slightly
from (3.14) to (3.16).  Make the shift
(3.22) in (3.14).  Then note that the
last term of (3.16) is 
\be%3.24
\frac{N_f}{2} \:
\L^{2r+2-N_f} \: \frac{x^{N_f-1}}{y^3} \; .
\ee
For $r+2 \leq N_f \leq 2r+1$ the period integral
$\int dx \: \frac{x^{N_f-1}}{y^3}$ is in the basic
range of $\O^{(1/2)}_m$.  Therefore, for these
flavors the steps described in (3.17) to (3.20) are
applicable to (3.24).  This again requires a symbolic
computer calculation.

For $N_f = r+1$, the situation is more complicated
as $\int x^r/y^3$ is not in the basic range of
$\O^{(1/2)}_m$.  However, this is exactly the problem
dealt with in section 3.3 of ref. \cite{023}.  [Note that
$n=r+1$ in that reference.]  One then follows the
scheme described there to reduce the $2g \times
(2g+1)$ $M$ matrix to the required $2g \times 2g$
$M$ matrix.  Only then will the reduced $M$ matrix
be suitable for application of (3.17) to (3.20).
All these steps can be carried out by symbolic
computer calculations, as in ref. \cite{023}.

The last PF equation is obtained from the 
generalization of ref. \cite{028}
\begin{eqnarray}%3.25
D & = & \l + \sum^{r+1}_{j=2} \, j (j-2)
\bar{u}_j \pa_j \, \l \nonumber \\
&& + \; \sum^{r+1}_{i,j=2} \: i j \bar{u}_i
\bar{u}_j \, \pa_i \pa_j \, \l
\nonumber \\
&& + (r +1 - \frac{N_f}{2})^2 \,
\L^{2r+2-N_f} \, \frac{x^{N_f}}{y^3} \:
p(x) \, dx
\end{eqnarray}
where $\bar{u}_i$ is defined by (3.22).  
Using (3.3), one must relate
\be%3.26
\frac{x^{N_f}}{y^3} \: p(x) = 
\frac{x^{N_{f}+r+1}}{y^3} - \sum^{r+1}_{i=2} \;
\bar{u}_i \:
\frac{x^{N_{f}+r+1-i}}{y^3}
\ee
to $\l$.  The terms of $\int \, \frac{x^m}{y^3}$ for
$m > 2r$ are outside the basic range.  These must be
brought back to the basic range using section 2
of ref. \cite{023}, or the methods of refs. \cite{010} and \cite{027}.
Once $\O^{(1/2)}_m$ lies in the basic range, then
one applies (3.17)--(3.20) to relate the results
to $\l$.

For $N_f = 2r+1$ the shift (3.22) is not valid, and
$u_i$, not $\bar{u}_i$, appears in (3.25).  Finally
the last term in (3.25) is now
\begin{eqnarray}%3.27
\frac{\L}{4}\frac{x^{2r+1}}{y^3} \: p(x) - \frac{\L}{8}\frac{x^{r}}{y} 
-\frac{\L^2}{16}\frac{x^{3r+1}}{y^3} 
- \frac{\L}{8}\,\sum^{r+1}_{i=2} \:
i\,u_i\, \:
\frac{x^{2r+1-i}}{y^3} \: p(x)
\end{eqnarray}
The method outlined in the previous paragraph must be
followed to relate (3.27) to $\l$.

Note that for $N_f \geq r+1$, only $r-2$ PF
equations have a generic presentation, while for
$1 \leq N_f < r+1$, there are $r-1$ generic
equations.  This is to be contrasted with 
$N_f = 0$, where all $r$ PF equations are generic.
Furthermore, we observe that the computer calculations are
only required for the $\L$-dependent term.

\ul{$B_r: \; 1 \leq N_f < (r-1)$}

The SW differential satisfies (2.21) to (2.23).  
Equation (2.24) is replaced by
\be%3.28
\frac{d}{dx} \, \left( \frac{x^{2n+1}}{y} 
\right) = \frac{(2n-N_f)x^{2n}}{y} - 
\frac{[x^{2n+1} p^\prime - (N_f + 1) x^{2n} p]}
{y^3} \; p(x)
\ee
which implies
\begin{eqnarray}%3.29
{\cal L}_n & = & (2n-N_f) \pa_{2r-2n} - 
(2r - 1 - N_f)
\pa_2 \, \pa_{2r-2n-2}\nonumber \\
&& + \; \sum^r_{j=1} (2r - 2j -1-N_f) u_{2j} \, 
\pa_{2j} \, \pa_{2r-2n} 
\hspace{.5in} (n=0 \; \mbox{\rm to} \; r-2)
\end{eqnarray}

Equations (2.26) and (2.27) trivially generalize
to give
\begin{eqnarray}%3.30
{\cal L}_{r-1} & = &
1 + 4 \: \sum^r_{j=1} \, j(j-1) u_{2j}\pa_{2j}
\nonumber \\
& + & 4 \: \sum^r_{i,j=1} \,i j u_{2i} u_{2j}
\pa_{2i} \pa_{2j}\nonumber \\
& - & (2r-1-N_f)^2 \L^{4r-2-2N_f}
\pa_{2r}\pa_{2r-2-2N_f}
\end{eqnarray}

\ul{$B_r: \; (r-1) \leq N_f \leq 2r-2$}

Equations (2.21)--(2.23) and (3.29) are 
applicable.  One may use equations (3.6) and (2.6) 
to shift the modulus
\be%3.31
u_{2i}  \rightarrow  u_{2i} - \L^{4r-2-2N_f} = 
\bar{u}_{2i}
\ee
where $2i = 4r - 2N_f -2$.  With this redefinition, 
(3.29) is applicable, but with a (single) shifted 
modulus.  That is,
\begin{eqnarray}%3.32
{\cal L}_n & = &(2n-N_f) \pa_{2r-2n}-(2r-1-N_f)\pa_2
\pa_{2r-2n-2}\nonumber \\
&& + \;\sum^r_{\stackrel{\textstyle j\neq i}
{\textstyle j=1}}
 \; (2r-2j-1-N_f) u_{2j} \pa_{2j} \pa_{2r-2n} 
\nonumber \\
&& + \; (2r-2i-1-N_f) (u_{2i} - \L^{4r-2-2N_f}) 
u_{2i}
\pa_{2i} \pa_{2r-2n}
\end{eqnarray}
where $i=2r-N_f-1$ and $n=0$ to $r-2$.

To obtain the last PF equation note that (3.30)
is applicable for $r-1 \leq N_f \leq 2r-3$, 
but with the substitution (3.31).

It is more complicated to obtain the last PF
equation for $N_f = 2r-2$.  In that case one
has from the analogue of (3.30)
\begin{eqnarray}%3.33
{\cal L}_{r-1}\, \l \; = \; \l & + &
4 \sum^r_{j=1} \: j (j-1) \bar{u}_{2j} \,
\pa_{2j} \, \l \nonumber \\
& + & 
4 \sum^r_{i,j=1} \:i j  \bar{u}_{2i}  
\bar{u}_{2j} \,
\pa_{2i} \pa_{2j}\, \l \nonumber \\
& + & \L^{2r} \: \frac{x^{4r-2}}{y^3} \:
p(x) \, dx \; ,
\end{eqnarray}
where $\bar{u}_{2j} = u_{2j}$ for $j\neq 1$,
and $u_2$ shifted by (3.31).  One cannot use
(2.21) for the last term, as this would not
involve allowed modular derivatives.  Since
$p(x) = x^{2r} - \sum^r_{i=1} \bar{u}_{2i}
x^{2r-2i}$, the period $\O^{(1/2)}_m$ is outside
the basic range.  One uses the recursion relations
of ref. \cite{023} to bring these periods back to the
basic range.  Then one uses (3.17)--(3.20) to
relate the last term of (3.33) to $\l$.  
Although there is no generic solution, note that
the computer calculation is restricted to the
$\L$-dependent part of (3.33).

\ul{$C_r: \; 2 \leq N_f \leq 2r+1$}

Equations (2.28) and (2.29) apply, together with
\begin{eqnarray}%3.34
\frac{d}{dx} \left( \frac{x^{2n+2}}{y} \right) 
& = & 
\frac{(2n+3-N_f)x^{2n+1}}{y} \nonumber \\
& - & \frac{[x^{2n+2} p^\prime + (1-N_f)
x^{2n+1}p]}{y^3} \; p(x)
\end{eqnarray}
which implies
\begin{eqnarray}%3.35
{\cal L}_n & = & (2n+3-N_f) \pa_{2r-2n} - 
(2r +2-N_f) \pa_2 \pa_{2r-2n-2}\nonumber \\
&& + \; \sum^r_{j=1} (2r + 2 -N_f-2j) u_{2j} 
\pa_{2j}\pa_{2r-2n}
\hspace{.5in} (n=0 \; \mbox{\rm to} \; r-2)
\end{eqnarray}

Using the strategy of ref. \cite{028}
\begin{eqnarray}%3.36
D & = & - (2r+2 - N_f) d
\left( \frac{x^{2r+2}}{y}\right) \nonumber \\
& + &  \sum^r_{j=1} (2r + 2 - N_f + 2j)
u_{2j} d \left( 
\frac{x^{2r+2-2j}}{y} \right)
\end{eqnarray}
leads to the equation
\begin{eqnarray}%3.37
{\cal L}_{r-1} & = &
1 + 4 \: \sum^r_{j=1} \, j(j-1) u_{2j}\pa_{2j}
\nonumber \\
& + & 4 \: \sum^r_{i,j=1} \,i j u_{2i} u_{2j}
\pa_{2i} \pa_{2j}\nonumber \\
& - & (2r+2-N_f)^2 \L^{4r+4-2N_f}
\pa_{2r} \pa_{2r+4-2N_f} \; ,
\end{eqnarray}
valid for $2 \leq N_f \leq 2r$.

{}For $N_f = 2r+1$, we have instead of (3.37)
\begin{eqnarray}%3.38
{\cal L}_{r-1} \l & = &
1 + 4 \: \sum^r_{j=1} \, j(j-1) u_{2j}\pa_{2j}\l
\nonumber \\
& + & 4 \: \sum^r_{i,j=1} \,i j u_{2i} u_{2j}
\pa_{2i} \pa_{2j} \l \nonumber \\
& + & \L^2 \: \frac{x^{4r}}{y^3} \: p(x) \, dx
\end{eqnarray}
where $p(x) = x^{2r+1} - \sum^r_{i=1} \,
u_{2i} \, x^{2r-2i+1}$.  Since
\be%3.39
\frac{x^{4r}p(x)}{y^3} = \frac{x^{6r+1}}{y^3} \:
- \: \sum^r_{i=1} \: u_{2i} \:
\frac{x^{6r-2i+1}}{y^3}
\ee
lies outside the basic range, the period
$\O^{(1/2)}_m$ must be brought back to the basic
range by the relations of ref. \cite{023}.  Then (3.17)
to (3.20) will relate the last term of (3.38) to
$\l$.  Once again the symbolic computer 
computation is only required for the quantum
term in (3.38).

\ul{$C_r: \;N_f=1$}

Equations (3.28) and (3.29) are applicable. Note 
that (3.37) does not apply to this case,
as the last term would not involve legitimate
derivatives with respect to moduli. Instead of
(3.37) we have
\begin{eqnarray}%3.40
{\cal L}_{r-1} \: \l & = &
\l + 4 \sum^r_{j=1} j(j-1) u_{2j}\pa_{2j} \: \l
\nonumber \\
& + & 4 \sum^r_{i,j=1} ij \, u_{2i} u_{2j}
\pa_{2i}\pa_{2j} \; \l
\nonumber \\
& + & (2r+1)^2 \, \L^{4r+2} \, \frac{p(x)}{y^3}
\: dx
\end{eqnarray}
where $p(x)$ is given by (2.6) and (3.7).  Thus
\be%3.41
\frac{p(x)}{y^3} = \frac{x^{2r+1}}{y^3} \: - \:
\sum^r_{j=1} \: u_{2j} \: 
\frac{x^{2r+1-2j}}{y^3}\; .
\ee
Using the definition (3.17), we consider the
periods required by (3.41), {\it i.e.},
$$
\O^{(1/2)}_{2r+1-2j} \; , \hspace{.3in}
(j = 0~~\makebox{\rm to}~~r)
$$
where the periods are ``odd" in the terminology
of ref. \cite{023}.   Equation (3.18) relates periods
$\O^{(1/2)}_{m^\prime}$ to periods 
$\O^{(-1/2)}_m$.  The periods 
$\O^{(-1/2)}_{2m+1}$, with $m=0$ to $r-1$ can
be related directly to $\l$, by means of (2.28).
That is
\be%3.42
\O^{(-1/2)}_{2m+1} = -{\rm i}\,\Gamma(1/2)\,\pa_{2r-2m}\,
\int dx \: \l \hspace{.4in}
\makebox{\rm for}~~m = 0~~\makebox{\rm to}~~r-1 \; .
\ee
However, for $m=r$ to $2r-1$ one must use the
analogue of (3.20).  Then the periods on the 
right side of (3.20) are expressed in terms of
$\l$ by means of (3.42).  Thus (3.40) to (3.42),
with (3.18) and (3.20), provides us with the final
second order PF equation for $N_f=1$.  The
computer calculations are only needed for the
$\L$-dependent part of (3.40).

\ul{$D_r: \; 1 \leq N_f < (r-2)$}

The SW differentials satisfy (2.34) and (2.35). We also have
\begin{eqnarray}%3.43
\frac{d}{dx} \left( \frac{x^{2n+1}}{y} \right) 
& = &
\frac{(2n-1-N_f)x^{2n}}{y} \nonumber \\
&& - \; \frac{[x^{2n+1} p^\prime - (2 + N_f) 
x^{2n}p]}{y^3} \: p(x)\; .
\end{eqnarray}
Hence
\begin{eqnarray}%3.44
{\cal L}_n & = & (2n-1-N_f) \pa_{2r-2n} - 
(2r-2-N_f) \pa_2\pa_{2r-2n-2}
\nonumber \\
&& + \; \sum^r_{j=1} (2r-2-2j-N_f) u_{2j} \, 
\pa_{2j} \pa_{2r-2n}
\hspace{.5in} (n=0 \; \mbox{\rm to} \; r-2)
\end{eqnarray}
Equations  (2.38) and (2.39) generalize to
\begin{eqnarray}%3.45
{\cal L}_{r-1} & = &
1 + 4 \: \sum^r_{j=1} \, j(j-1) u_{2j}\pa_{2j}
\nonumber \\
& + & 4 \: \sum^r_{i,j=1} \,i j u_{2i} u_{2j}
\pa_{2i} \pa_{2j}\nonumber \\
& - & \left[ (2r-2-N_f)^2 \L^{4r-4-2N_f}
\pa_{2k} \pa_{2l} \right]_{k+l =2r-2-N_f} \; .
\end{eqnarray}

\ul{$D_r: \; (r-2) \leq N_f \leq 2r-3$}

{}From (2.6) and (3.11) we can shift
\be%3.46
u_{2i} \rightarrow u_{2i} - \L^{4r-4-2N_f} 
= \bar{u}_{2i}
\ee
where $2i = 4r - 2N_f - 4$.  With this 
redefinition, there are $r-1$ PF equations 
identical to (3.44), but with the single shifted 
modulus (3.46).

When $r-2 \leq N_f \leq 2r-4$ the last PF equation 
is obtained from (3.45) with the substitution (3.46).

$N_f =2r-3$ must be considered separately.  Instead
of (3.45), one has
\begin{eqnarray}%3.47
{\cal L}_{r-1} \l & = &
\l + 4 
\: \sum^r_{j=1} \, j(j-1) \bar{u}_{2j}\pa_{2j} \l
\nonumber \\
& + & 4 \: \sum^r_{i,j=1} \,i j \bar{u}_{2i} 
\bar{u}_{2j}\pa_{2i} \pa_{2j} \l \nonumber \\
& + &  \L^2 \: \frac{x^{4r-2}}{y^3} \:
p(x) \, dx
\end{eqnarray}
with $\bar{u}_{2j} = u_{2j}$ for $j \neq 1$, and
$u_2$ given by (3.46).  Then with (3.11) for $p(x)$
one uses the strategy outlined in (3.33) {\it ff}.
It is obvious once again that the computer
calculation is only required for the $\L$-dependent
part of (3.47).

{\bf 4. ~Concluding Remarks}

We have provided a systematic presentation
of the PF equations for $N$=2 theories with massless
multiplets presently available.
For $N_f =0$ all $r$ equations can be put in
generic form.  When $N_f \neq 0$ there are at least
$r-2$ equations which can be formulated generically, with only the
quantum part requiring a case-by-case analysis. Although these symbolic
computer calculations are exhibited here, the structure of the remaining
1 or 2 equations is quite explicit for the classical part.
It is our belief that
this compilation of PF equations will facilitate 
further understanding of the
Coulomb phase of these theories in the strong-coupling
regions.

We wish to thank S. Naculich and H. Rhedin for
helpful conversations.

After this paper was completed we received the paper \cite{032} which has 
considerable overlap with our work.


\begin{thebibliography}{99}
\bibitem{001}
N. Seiberg and E. Witten, 
\np {\bf B426} (1994) 19, {\tt (hep-th/9407087)}; 
\np {\bf B431} (1994) 484  {\tt (hep-th/9408099)}.
\bibitem{002}
A. Klemm, W. Lerche, S. Theisen and S. Yanckielowicz
\pl {\bf B344} (1995) 169 {\tt (hep-th/9411048)}.
\bibitem{003}
P.C. Argyres  and A.E. Faraggi, \prl {\bf 74} (1995) 3931 
{\tt (hep-th/9411057)}.
\bibitem{004}
A. Brandhuber and K. Landsteiner, \pl {\bf B358} (1995) 73  
{\tt (hep-th/9507008)}.
\bibitem{005}
A. Hanany and Y. Oz, \np {\bf B452} (1995) 283
 {\tt (hep-th/9505075)}.
\bibitem{006}
A. Hanany, \np {\bf B466} (1996) 85 {\tt (hep-th/9509176)}.
\bibitem{007}
P.C. Argyres and A. Shapere, \np {\bf B461} (1996) 437
{\tt (hep-th/9509175)}.
\bibitem{008}
M.R. Abolhasani, M. Alishahiha and A.M. Ghezelbash, 
{\tt (hep-th/9606043)}.
\bibitem{009}
U.H. Danielsson and B. Sundborg, \pl {\bf 358} (1995) 73 {\tt
 (hep-th/9504102)}.
\bibitem{010}
A. Klemm, W. Lerche and S. Theisen, \ijmp {\bf A11} (1996)
1929 {\tt (hep-th/9505150)}.
\bibitem{011}
M. Douglas and S.H. Shenker, \np {\bf B447} (1995) 271  
{\tt (hep-th/9503163)}.
\bibitem{012}
P.C. Argyres and M. Douglas, \np {\bf B448} (1995) 166
{\tt (hep-th/9505062)}.
\bibitem{013}
P.C. Argyres, M.R. Plesser and A.D. Shapere, \prl {\bf 75}
(1995) 1699 
{\tt (hep-th/99505100)}.
\bibitem{014}
T. Eguchi, K. Hori, K. Ito and S. -K. Yang 
\np {\bf B471} (1996) 430 {\tt (hep-th/9603002)}.
\bibitem{015}
U.H. Danielsson and B. Sundborg, \pl {\bf B370} (1996) 83
 {\tt (hep-th/9511180)}.
\bibitem{016}
P.C. Argyres, M.R. Plesser and N. Seiberg, 
{\tt (hep-th/9603042)}.
\bibitem{017}
J.A. Minahan and D. Nemeshansky, \np {\bf B464} 
(1996) 3 {\tt (hep-th/9507032)}.
\bibitem{018}
E. D'Hoker, I. Krichever, and D. Phong {\tt (hep-th/9609041)};
{\tt (hep-th/9609145)}; E. D'Hoker and D. Phong
{\tt (hep-th/9701055)}.
\bibitem{019}
M. Matone, Phys. Lett. {\bf B357} (1995) 342; 
G. Bonelli and M. Matone, \prl {\bf 76} (1996) 4107; 
G. Bonelli and M. Matone, ({\tt hep-th/9605090})
\pr{\bf 77} (1996) 4712;  
G. Bonelli, M. Matone and M. Tonin, ({\tt hep-th/9610026}).
\bibitem{020}
L. \'Alvarez-Gaum\'{e}, J. Distler, and M. Mari\~{n}o,
Int. J. Mod. Phys. {\bf A11} (1996) 4745 
({\tt hep-th/9604004});
L. \'Alvarez-Gaum\'{e} and M. Mari\~{n}o,
Int. J. Mod. Phys. {\bf A12} (1997) 975 
({\tt hep-th/9606191}); ({\tt hep-th/9606168});
L. \'Alvarez-Gaum\'{e} and S. F. Hassan, ({\tt hep-th/9701069});
L. \'Alvarez-Gaum\'{e}, M. Mari\~{n}o, and
F. Zamora ({\tt hep-th/9703072}); N. Evans, S.D.H. Hsu,
and M. Schwetz ({\tt hep-th/9608135}).
\bibitem{021}
K. Ito and S.-K. Yang, \pr {\bf D53} (1996) 2213 
{\tt (hep-th/9603073)};  \pl {\bf B366} (1996) 165 
{\tt (hep-th/9507144)}. 
\bibitem{022}
K. Ito and N. Sasakura, {\tt  (hep-th/9608054)}
\np {\bf B484} (1997) 141;
K. Ito, {\tt (hep-th/9703180)}. 
\bibitem{023}
J.M. Isidro, A. Mukherjee, J.P. Nunes, and H.J. Schnitzer,
\np {\bf B} (to be published) ({\tt hep-th/9609116)}
\bibitem{024}
Y. Ohta, {\tt (hep-th/9604051)}; {\tt (hep-th/9604059)}.
\bibitem{025}
S. Ryang, \pl {\bf B365} (1996) 113  
{\tt (hep-th/9508163)}.
\bibitem{026}
A. Bilal, {\tt (hep-th/9601007)}; 
A. Bilal and F. Ferrari, {\tt (hep-th/9605101)}
\np {\bf B480} (1996) 589;
A. Bilal and F. Ferrari, {\tt (hep-th/9602082)}
\np {\bf B469} (1996) 337; 
{\tt (hep-th/9606111)};
F. Ferrari, {\tt (hep-th/9702166)}; J. Schulze and N. P. Warner, 
{\tt (hep-th/9702012)}; J. M. Rabin, {\tt (hep-th/9703145)}.
\bibitem{027}
H. Ewen, K. Foerger and S. Theisen,  
\np {\bf B485} (1997) 63
{\tt (hep-th/9609062)}; H. Ewen and K. Foerger, 
({\tt hep-th/9610049}).
\bibitem{028}
M. Alishahiha, ({\tt hep-th/9609157}).
\bibitem{029}
A. Mukherjee, PhD Thesis, University of Cambridge, 1995.
\bibitem{030}
M. Slater, ({\tt hep-th/9601170}); N. Dorey, V.
Khoze, and M. Mattis, ({\tt hep-th/9612231});
\pr {\bf D54} (1996) 2921, ({\tt hep-th/9603136});
\pl {\bf B388} (1996) 324, ({\tt hep-th/9607066});
\pr {\bf D54} (1996) 7832, ({\tt hep-th/9607202});
\pl {\bf B390} (1997) 205, ({\tt hep-th/9606199});
K. Ito and N. Sasakura, \pl {\bf B382} (1996) 95,
({\tt hep-th/9602073}); ({\tt hep-th/9609104});
H. Aoyama, {\it et al.}, \pl {\bf B388} (1996) 331,
({\tt hep-th/9607076}); T. Harano and M. Sato,
({\tt hep-th/9608060}).
\bibitem{031}
J.M. Isidro, A. Mukherjee, J.P. Nunes, and H.J. Schnitzer,
in preparation.
\bibitem{032}
M. Alishahiha, ({\tt hep-th/9703186}).
\end{thebibliography}
\end{document}